# Robust population transfer by finite chirping method in a two-level system

Fatemeh Ahmadi Nouri [1], Mehdi Hosseini,[1a] and Farrokh Sarreshtedari [2]

*[1]Department of physics, Shiraz University of Technology, Shiraz, 313-71555, Iran*

*[2]Magnetic Resonance Research Laboratory, Department of Physics, University of Tehran,*

*Tehran 143-9955961, Iran*

Considering a two-level quantum system, we have proposed and represented a new approach for robust population transfer. In this scheme, the laser frequency has been swept in a finite time interval which simplifies the experimental limitations of the population transfer process. It is shown that using the Jaynes-Cummings model and engineering the coupling strength, the frequency sweeping range and its time interval, it is possible to achieve a robust, stable and full population transfer.

PACS numbers: 42.65.Re, 32.80.Xx, 32.80.-t

Keywords: population transfer, two-level system, chirped laser

## I. INTRODUCTION

Many different atomic physic and chemistry experiments are engaged with the population transfer between two level or multi-level quantum systems including atomic and molecular spectroscopy [1,2], atomic clocks [3,4], interferometry [5], magnetic resonance [6], quantum information processing [7–9], chemical reactions [10,11], collision dynamics [12] and laser cooling [13]. Because of that, the population transfer plays a significant role in quantum engineering and working towards developing robust and efficient transferring techniques has been always an attractive research field.

---

[a] hosseini@sutech.ac.ir

A suitable technique requires to ensure essential requirements [14] such as: (I) Efficiency and selectivity: the population requires to be entirely transferred to a selected state. (II) Robustness: the population transfer should be insensitive to small variations of the system parameters.

In this regard, one of the well-known techniques of the population transfer is the half-cycle Rabi oscillation which is called "$\pi$ *pulse*" [15]. In this method, the pulse area or flip angle should be adjusted to $\pi$ radian. Although this approach is one of the fast and complete population transfer in the individual atom it has two drawbacks: (I) It is highly sensitive to the fluctuations of different system parameters and consequently, it is far from robustness  [15]. (II) In an ensemble of dipole moments, the pulse area is tuned to the $\pi$ value for only a group of the dipole moment [16]. In contrast, "*adiabatic passage*" methods fulfil the essential population transferring requirements. Adiabatic passage approaches present high robustness, efficiency and selectivity versus the variation of the parameters.  Based on the adiabatic dynamics, the system parameters change very slowly to preserve an eigenstate of the system [17,18]. One of the prominent techniques of the adiabatic population transfer is the Stimulated Raman Adiabatic Passage (STIRAP) [19]. In this approach, the system population is transferred by two delayed counterintuitive-ordered pulses under the resonance and adiabatic condition. The first pulse (Stokes) couples the intermediate (dark) state to the excited state. In the following, the ground state is connected by a second pulse (pump) to the intermediate state. This technique has been widely investigated both theoretically and experimentally [19]. However, this approach requires precise controlling of the Stokes and pump pulses. Furthermore  the inhomogeneous broadening make problems for the application of STIRAP method and reduces its efficiency  [20]. This is while, using chirped laser pulses in adiabatic regime,  complete and efficient population transfer can be achieved  [21–25]. In this approach, the time-dependent laser frequency is slowly swept from far below resonance to far above resonance or vice versa. This approach is insensitive to the pulse area and to the exact resonance frequency thus this scheme can be used for an ensemble of atoms [26]. Therefore this method is simpler than the STIRAP [22]. However, one of the problems in this method is the experimental implementation of the infinite frequency chirping of the laser.

Hence, we have represented a new approach for robust and complete population transfer with minimum chirping of the laser frequency. This approach significantly simplify the experimental realization of the method.

## II. THEORETICAL MODEL

In this work, the conventional Jaynes-Cummings (JC) model have been used to investigate the transition probabilities of a two-level system which interacts with an external electric field [27].

$$H = \hbar\omega_L(a^\dagger a + \frac{1}{2}) - \frac{1}{2}\hbar\omega_0\sigma_z + \hbar\omega_1(a\sigma_+ + a^\dagger\sigma_-) \tag{1}$$

The JC Hamiltonian (Eq. 1) includes three terms corresponding to the field, atom and atom-field interaction, where $\tilde{\omega}_L$ is the laser frequency, $\tilde{\omega}_0$ is the atomic transition frequency and $\tilde{\omega}_1$ describes the coupling strength between atom and field. Moreover, $a$, $a^\dagger$ are annihilation and creation operators respectively, $\hbar$ is Planck constant and $\sigma_-$, $\sigma_+$, $\sigma_z$ are Pauli spin matrices. In this definition the coupling strength is determined by the atomic transition dipole moment and the laser field which is proportional to Rabi frequency as follows [28]:

$$\Omega = \sqrt{(\omega_L - \omega_0)^2 + (2\omega_1)^2} \tag{2}$$

By applying the JC Hamiltonian to eigenstates $|g, n\rangle$ (ground state, n photons) and $|e, n\rangle$ (excited state, $n$ photons) and by solving the time-dependent Schrödinger equation the results are obtained and discussed [29]. In this calculation, $n$ is equal to one.

It should be noted that for simplicity all the parameters are dimensionless using the following relations:

$$\tilde{\omega}_0 = 1, \tilde{t} = t\,\omega_0, \tilde{\alpha} = \frac{\alpha}{\omega_0^2}, \tilde{\omega}_1 = \frac{\omega_1}{\omega_0}, \tilde{\omega}_{L0} = \frac{\omega_{L0}}{\omega_0} \qquad (3)$$

## III. RESULTS AND DISCUSSION

Here, solving the time-dependent Schrödinger equation by Runge–Kutta method, the transition probabilities are calculated.

In this research, the frequency chirping of the laser is applied for a finite time interval in such a way that the laser frequency remains constant before and after the chirping interval. Then it is attempted to reach to a robust population transfer with the minimum chirping interval (sweep). This approach helps to simplify the experimental implement the system population transfer. Here, the angular frequency of the laser is given by:

$$\tilde{\omega}_L = \tilde{\alpha}\tilde{t} + \tilde{\omega}_{L0} \qquad (4)$$

Where $\omega_{L0}$ is the center frequency, $\alpha$ is the chirping constant.

In Fig. 1 the transition probability versus time are depicted. The time interval of the frequency chirping is $-100 < \tilde{t} < 100$ while the laser frequency remains constant before and after this interval. In the inset of these figures laser frequency sweeping range ($\Delta\tilde{\omega}_L = \tilde{\omega}_{L2} - \tilde{\omega}_{L1}$) versus time are illustrated. The parts (a) and (b) represent that by increasing the laser frequency sweeping range, the final probability is reduced. The parts (a) and (c), by increasing the initial laser frequency ($\tilde{\omega}_{L1}$) the final probability remains constant while the amplitude of initial oscillations is increased. Although increasing coupling strength in the parts (b) and (d), the final probability is increased but the amplitude of initial oscillations is increased. Our findings indicate that increasing initial laser frequency, the number (frequency) of initial oscillations is reduced whereas increasing coupling strength, the number of initial oscillations is increased.

Since the final probability is related to the initial condition; changing amplitude of initial oscillations, the final probability is changed. In order to investigate the population transfer and its changes MFPG is introduced which express Mean values in the 10% Final Probability of Ground state. Fig. 2 shows the MFPG versus laser frequency sweeping range ($\Delta\tilde{\omega}_L$) and sweeping time interval ($\Delta\tilde{t}$) for different coupling strength and different initial laser frequency while the resonance frequency equals to 1 ($\tilde{\omega}_L$=1). It should be noted that for each $\Delta\tilde{\omega}_L$ and $\Delta\tilde{t}$ in this figure, the laser frequency has been changed from $\tilde{\omega}_{L1}$ to $\tilde{\omega}_{L2}$ in the interval of $\Delta\tilde{t}$. The blue region illustrates that the population is transferred to the excited state. This figure reveals that the transition probability is increased by increasing the $\Delta\tilde{t}$ and reducing the $\Delta\tilde{\omega}_L$. This means that the population is more efficiently transmitted for longer time and less frequency sweep. Fig. 2 (a, b, c) shows that no population transfer take place for $\Delta\tilde{\omega}_L$ <1, when the laser frequency does not pass the resonance frequency. By increasing the initial laser frequency, this border is reached to zero (Fig. 2 (g, h, i)). It corresponds to the fact that the laser frequency no longer would be in resonance with the resonance frequency ($\tilde{\omega}_L$=1). Furthermore, it is evident that by increasing the coupling strength, the region of the transition is increased (Fig. 2 (b, e)). This is while by further increasing of the coupling strength, the complete population transfer does not happen (Fig. 2 (c, f, i)). It can be inferred that by increasing the coupling strength, the avoided crossing would be increased and the population transfer would be improved [30] whereas by further increasing of the coupling strength the transition probability would be reduced [31]. In Fig.2 (b) the full transition regions with a high degree of stability and robustness is observed. By increasing the initial laser frequency, the robust area is decreased (Fig. 2 (e)). In this figure, an oscillating behavior is observed which the number of these oscillation has drastically increased by increasing the coupling strength. This is while increasing the initial laser frequency, the number of these oscillation decrease and their fluctuations amplitude increase. This is because while the initial laser frequency approaches to the resonance, the amplitude of initial oscillations is increased and reached to one at $\tilde{\omega}_L$=1. Since the final probability is related to the initial condition, accordingly, the fluctuation amplitude of MFPG is drastically increased and their oscillation number is decreased (as seen in Fig.1). Therefore, by adjusting laser parameters and coupling strength, it is possible to the complete and robust population transfer in the infinite frequency chirping as seen in this figure (Fig.1 (b, e)).

Moreover, in order to achieve stable population transfer, the amplitude of final probability oscillations should be minimized. For this reason, AFPG is introduced which is Amplitude of the Final Probability oscillations of Ground state (the value of (m-n) in Fig. 1(a)). In Fig. 3 the AFPG versus sweeping range ($\varDelta\tilde{\omega}_L$) and sweeping time interval ($\Delta \tilde{t}$) is depicted for different coupling strength and different initial laser frequency. Because, increasing the coupling strength between atom and laser is caused that the atom is more affected by the field, the amplitude of oscillations is increased and thus the AFPG is increased (Fig. 3 (b, d)). This figure also confirms that the maximum AFPG happens at the resonance ($\tilde{\omega}_L$=1) where this maximum of oscillation reach to zero by increasing the initial frequency (Fig. 3 (c, d)). Although, the Rabi oscillation oscillate between probabilities 0 and 1 at the resonance, the oscillation amplitude is decreased when it goes away from the resonance thus the AFPG is reduced. Therefore, the AFPG is optimal when the system is at the least coupling strength and away from the resonance which ensure that the system population would be in a steady state. Therefore, by simultaneous investigation MFPG and AFPG, In addition to the complete and robust population transfer, the stability of the final probability can also be guaranteed.

So far the effect of positive chirping frequency on population transfer was investigated. In Fig. 4 the complete population transfer is shown for positive and negative chirping of the laser frequency. In this figure, MFPG versus initial laser frequency ($\tilde{\omega}_{L1}$) and final laser frequency ($\tilde{\omega}_{L2}$) is depicted for different coupling strength and different time interval. For values of $\tilde{\omega}_{L1}$ and $\tilde{\omega}_{L2}$ which both of them are either less or more than one, no transition is seen. This is because that the laser frequency no longer pass the atomic transition frequency. For areas with the $\tilde{\omega}_{L1}$ less (more) than one, the regions of robust and complete transition are observed by increasing (decreasing) the $\tilde{\omega}_{L2}$. By further increasing (decreasing) of the $\tilde{\omega}_{L2}$, the transition probability is reduced and the oscillating behavior is increased (in accordance with Fig. 2). Our findings indicate that the transition probability eventually reaches to zero. In other words, the transition probability is reduced by increasing the positive (negative) chirping frequency [32]. In this figure, the robust area is increased by increasing the coupling strength (in agreement with Fig. 2 (b, e)). By further increasing of the coupling strength, the region of the complete transition is observed in higher final laser frequencies, which means that more chirping is needed for full population transfer (Fig. 4 (c, f)). This figure also confirms that by reducing the chirping time interval, the region of complete transition is decreased (Fig. 1 (d, e, f)). Indeed, by increasing the time interval and decreasing of the frequency sweeping range, more efficient population transfer with higher robustness would be occurred. this is while by adjusting the coupling strength, the value of the laser frequency sweeping range (chirping) and the time interval, it is possible to achieve the robust complete population transfer with the lowest frequency sweeping range in the lowest time interval (Fig. 4 (b)).

Furthermore, the AFPG is investigated for all cases. In Fig. 5, AFPG versus initial laser frequency and final laser frequency is depicted for different coupling strength. As mentioned before, the AFPG is maximized in the resonance mode because of the full Rabi oscillation. It is also evident that the AFPG is increased by increasing the coupling strength (Fig. 5 (b)) which is in accordance with Fig. 3. Therefore, by the appropriate selecting of the range and time interval of positive and negative laser frequency chirping, the complete, stable and robust population transfer can be achieved.

## IV. CONCLUSION

In this work, In order to facilitate the implementation of the population transfer for experimental purposes, the frequency has been chirped in a finite range. The effect of the chirping time interval, the value of the frequency sweeping and the coupling strength has been studied. Moreover, the stability of the final probability has been studied and it is shown that by appropriate setting of these parameters, it is possible to achieve the regions with a high robustness and stable full population transfer using the partially chirped lasers.

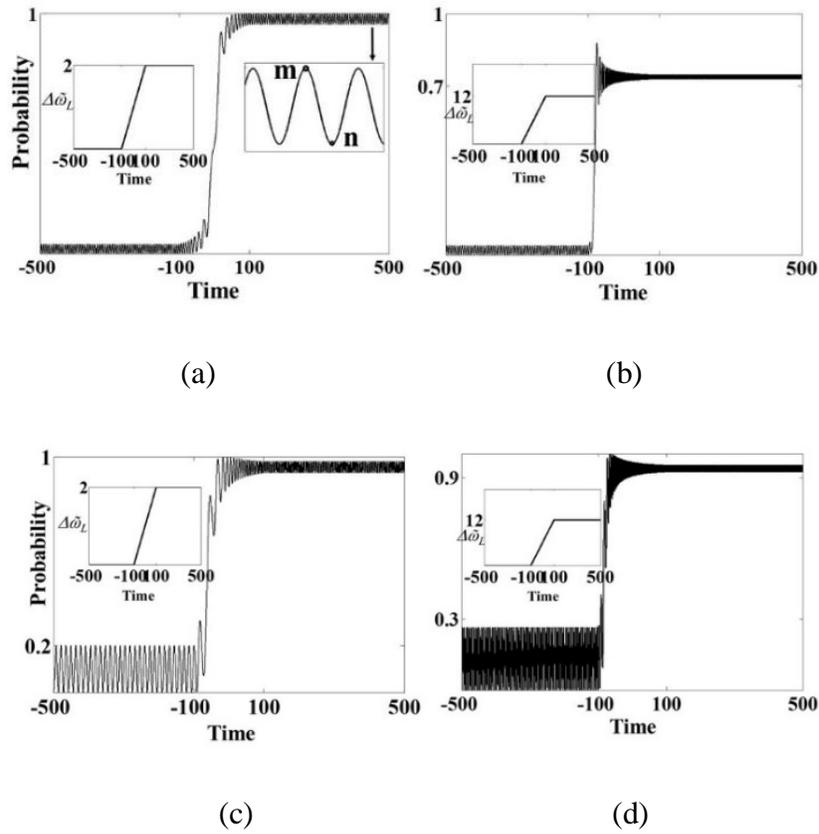

(a)                              (b)

(c)                              (d)

FIG.1. Transition probability versus time. The chirping time interval -$100 < \tilde{t} < 100$, (a) $\tilde{\omega}_I$=0.1, $\Delta\tilde{\omega}_L$=2, $\tilde{\omega}_{LI}$=0 (b) $\tilde{\omega}_I$=0.1, $\Delta\tilde{\omega}_L$=12, $\tilde{\omega}_{LI}$=0 (c) $\tilde{\omega}_I$=0.1, $\Delta\tilde{\omega}_L$=2, $\tilde{\omega}_{LI}$=0.6 (d) $\tilde{\omega}_I$=0.3, $\Delta\tilde{\omega}_L$=12, $\tilde{\omega}_{LI}$=0.

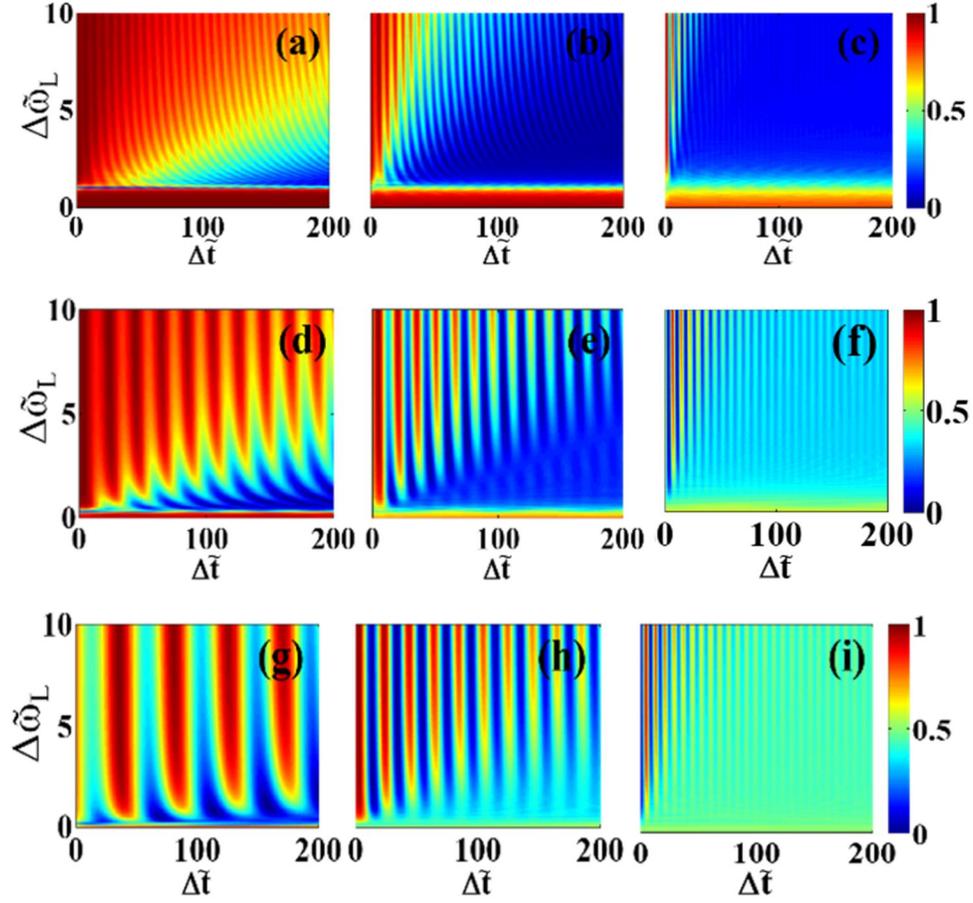

FIG.2. MFPG versus $\Delta\tilde{\omega}_L$ and $\Delta\tilde{t}$, (a) $\tilde{\omega}_1$=0.05, $\tilde{\omega}_{L1}$=0, (b) $\tilde{\omega}_1$=0.15, $\tilde{\omega}_{L1}$=0, (c) $\tilde{\omega}_1$=0.4, $\tilde{\omega}_{L1}$=0, (d) $\tilde{\omega}_1$=0.05, $\tilde{\omega}_{L1}$=0.7, (e) $\tilde{\omega}_1$=0.15, $\tilde{\omega}_{L1}$=0.7, (f) $\tilde{\omega}_1$=0.4, $\tilde{\omega}_{L1}$=0.7, (g) $\tilde{\omega}_1$=0.05, $\tilde{\omega}_{L1}$=0.9, (h) $\tilde{\omega}_1$=0.15, $\tilde{\omega}_{L1}$=0.9, (i) $\tilde{\omega}_1$=0.4, $\tilde{\omega}_{L1}$=0.9.

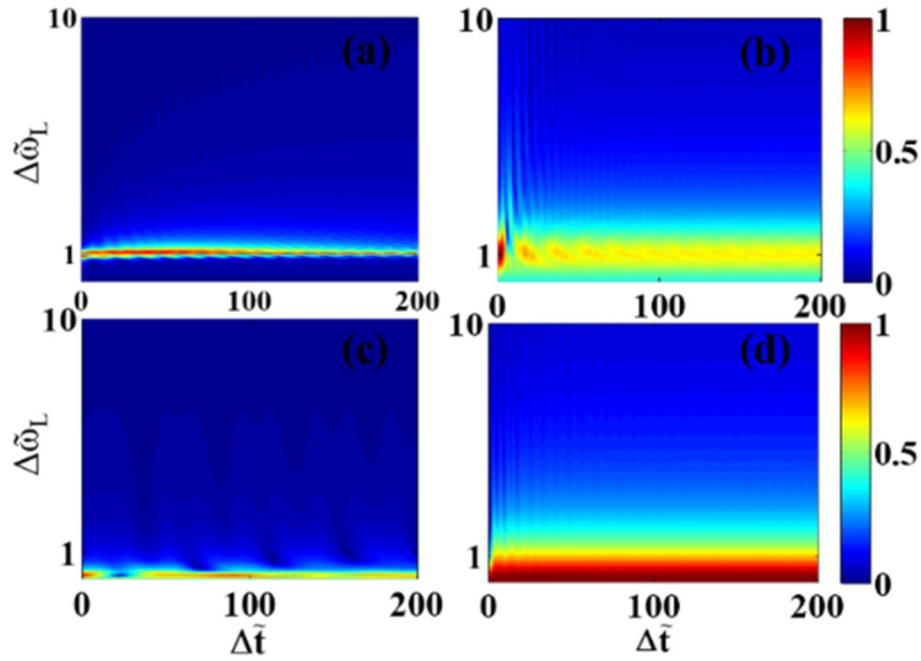

FIG.3. AFPG versus $\Delta\tilde{\omega}$ and $\Delta\tilde{t}$ , (a) $\tilde{\omega}_I$=0.05, $\tilde{\omega}_{LI}$=0, (b) $\tilde{\omega}_I$=0.4, $\tilde{\omega}_{LI}$=0, (c) $\tilde{\omega}_I$=0.05, $\tilde{\omega}_{LI}$=0.9, (d) $\tilde{\omega}_I$=0.4, $\tilde{\omega}_{LI}$=0.9.

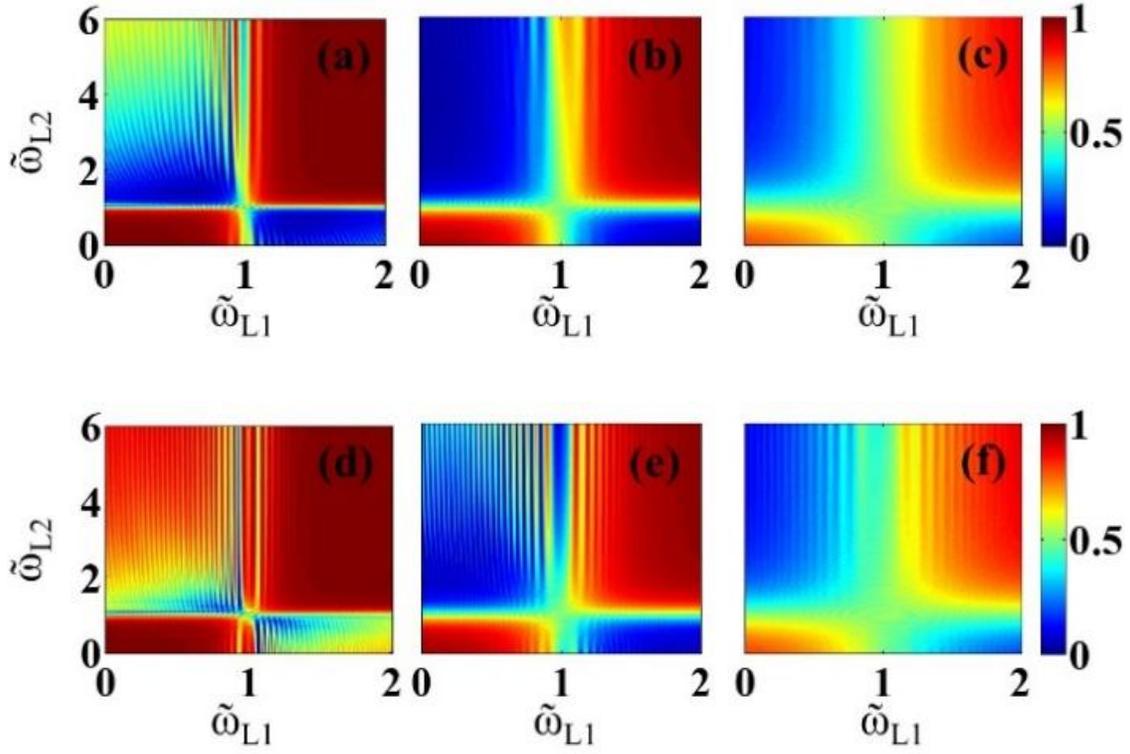

FIG.4. MFPG versus $\tilde{\omega}_{L0}$ and $\tilde{\omega}_{L1}$ (a) $\tilde{\omega}_{1}$=0.05, $\Delta\tilde{t}$ =200 (b) $\tilde{\omega}_{1}$=0.15, $\Delta\tilde{t}$ =200 (c) $\tilde{\omega}_{1}$=0.4, $\Delta\tilde{t}$ =200 (d) $\tilde{\omega}_{1}$=0.05, $\Delta\tilde{t}$ =50 (e) $\tilde{\omega}_{1}$=0.15, $\Delta\tilde{t}$ =50 (f) $\tilde{\omega}_{1}$=0.4, $\Delta\tilde{t}$ =50.

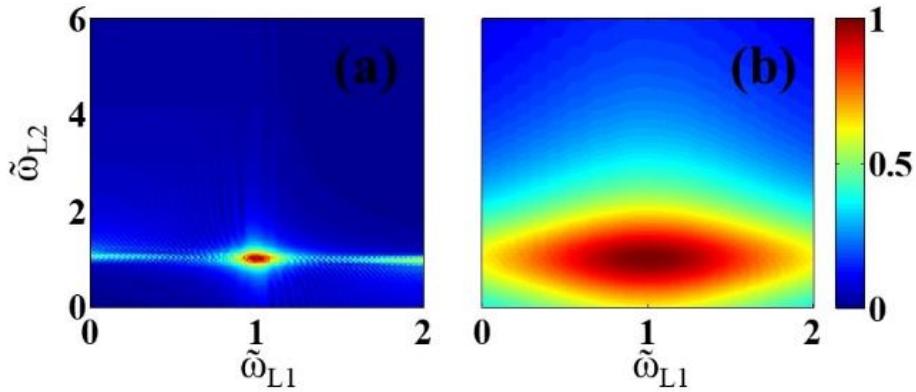

FIG.5. AFPG versus $\tilde{\omega}_{L1}$ and $\tilde{\omega}_{L2}$ (a) $\tilde{\omega}_{1}$=0.05, $\Delta\tilde{t}$ =200, (b) $\tilde{\omega}_{1}$=0.4, $\Delta\tilde{t}$ =200.